\documentclass[11pt,1p,review,number]{elsarticle}

\usepackage{fixltx2e} 
\usepackage[squaren,Gray]{SIunits}
\usepackage{natbib}
\biboptions{sort&compress}
\usepackage{custom-header}

\makeatletter
 \renewcommand\@journal{Acta Materialia}
\makeatother

\usepackage[pagewise,modulo]{lineno}

\begin{document}
\begin{frontmatter}

    \title{A new nanometer resolution method for probing densification ratio at nanoindentation sites in glass: unravelling discrepancies in the literature}
    \author[UR1]{J.-P. Guin}
    \author[UR1]{K. Han}
    \author[USMB]{L. Charleux}
    \author[UR1]{J.-C. Sangleboeuf}
    \author[UNSW]{M. Ferry}
    \author[UBS]{V. Keryvin\corref{cor2}}
    \address[UR1]{Univ. Rennes 1, UMR CNRS 6251, IPR, F-35042 Rennes, France}
    \address[USMB]{Univ. Savoie Mont-Blanc, EA 4114, SYMME, F-74000, France}
    \address[UNSW]{School of Materials Science and Engineering, The University of New South Wales, Sydney, NSW 2052, Australia}
    \address[UBS]{Univ. Bretagne Sud, UMR CNRS 6027, IRDL, F-56321 Lorient, France}

    \cortext[cor2]{Corresponding author: vincent.keryvin@univ-ubs.fr (Vincent Keryvin)}


    \begin{abstract}
        The region of permanent densification beneath a Berkovich indentation imprint in silica glass is investigated using a novel chemical dissolution technique.
        The use of the similitude regime in sharp indentation testing allows one to record reliable data with a good spatial resolution that makes it possible to deal with low loads (typically below $\unit{10}\milli\newton$) and, more importantly, crack-free imprints.
        The densified zone dissolves more quickly than the non densified regions.
        The analysis of the results, along the vertical axis, indicates that the densification zone is rather homogeneous with a steep transition to the non densified zone.
        The size of the densification zone, with respect to the initial free surface, is estimated to be around two times the maximum penetration depth of the instrumented indentation test.
        These findings are compared with those obtained by numerical simulations using different constitutive equations from the literature.
        A very good concordance between Raman spectroscopy and chemical probe results is found for imprints made with no or few cracking events during indentation testing.
    \end{abstract}


\end{frontmatter}

\section{Introduction}

The highest strength value measured for pristine silica glass fibres is 10 GPa at room temperature \citep{Proctor1967}, but the extreme sensitivity of silicate glasses to surface damage is often reported to be the reason for low strength values of few tens of $\mega\pascal$ in manufactured structural glass parts \citep{EN572}.
Although the negative impact of surface damage on the durability of silica glass has been studied for quite a long time \citep{Grenet1889}, a complete understanding of this effect is far from being realized.
An efficient way to scientifically study surface damage in glass is to create controlled sharp contact conditions between a pyramidal indenter having a well-defined geometry and a prepared glass surface.
Under such contact conditions, oxide glasses accommodate deformation both by elastic and permanent deformation mechanisms among which one can distinguish a volume conservative one, shear flow, and a non volume conservative one, densification (permanent mass density increase) \citep{Ernsberger1968, Peter1970, Hagan1980b, Yoshida2005, Rouxel2010, Yoshida2010}.
Their respective contributions depends strongly on the pressure and shear state as well as on the chemical composition of silicate glasses \citep{Kurkjian1995a}.
From purely hydrostatic compression tests it was shown that silica glass (respectively window glass) exhibits a threshold value of $\unit{10}\giga\pascal$ (resp. $\unit{8}\giga\pascal$) below which no densification is observed \citep{Bridgman1953, Cohen1961, Christiansen1962, Mackenzie1963a, Ji2006}.
Above this threshold, the value of the permanent densification ratio, relative to the
initial mass density, increases monotonically with applied pressure up to a saturation value of $21\%$ (resp. $6 \%$) at a pressure of $\unit{25}\giga\pascal$ \citep{Rouxel2008}.

Numerical simulations of the indentation process are reported in the literature using Finite-Element Modelling (FEM) \citep{Xin2000, Kermouche2008, Gadelrab2012, Bruns2017, Molnar2017a, Bruns2020, Barthel2020}, Discrete Element Modelling (DEM) \citep{Jebahi2013} or Molecular Dynamics (MD) simulations \citep{Kilymis2013a, Yuan2014}.
They are all able to describe the macroscopical mechanical response of the test, that is the force-displacement curve.
Meanwhile, they report different microscopic mechanical fields beneath the imprint.
There is a clear need of additional experimental information on what takes place beneath the imprint to discriminate among different models.

Micro Raman spectroscopy \citep{Rouxel2008a, Perriot2006, Hehlen2010, Kassir-Bodon2012, Deschamps2013, Bruns2020, Gerbig2020}, Brillouin spectroscopy \citep{Tran2012} or Small Angle X-ray Scattering \citep{Furhmann2020} were successfully used to either characterise permanently densified samples and their associated structural modifications or map out the size, shape and intensity of the densification ratio around and underneath residual indentation imprints \citep{Yoshida2005, Perriot2006,Bruns2020}.
Although these spectroscopic techniques generate invaluable information regarding structural changes, they suffer from a relatively low spatial resolution, which is of the order of a micrometer at best
\citep{Kassir-Bodon2012, Tran2012, Gerbig2020}.
Thus, to map out densification contrast under indentation imprints with sufficient spatial resolution, researchers resorted to increasing the size of the processed zone by the use of high indentation loads of at least $\unit{20}\newton$ \citep{Perriot2006, Kassir-Bodon2012, Tran2012} for silica glass or soda-lime-silica glass (the recent study of \citep{Bruns2020} used loads between $\unit{5}\newton$ and $\unit{10}\newton$ nevertheless).
However, as shown in \citep{Kassir-Bodon2012}, this level of loading results in massive fracturing of the zone underneath a Vickers indentation \citep{Hagan1979, Rouxel2001}.
Cracking events should be limited as much as possible using such methods as nano-indentation at loads
below $\unit{50}\milli\newton$ in silica glass \citep{Charleux2014b}.
An alternative technique to Raman or Brillouin spectroscopies, which relies on the increase in the rate of dissolution of silicate glasses with the level of densification, was developed by Niu \etal~\citep{Niu2012}.
By coupling congruent dissolution steps with atomic force microscopy (AFM) measurements, it was shown that measuring nanometer-sized changes in the shape of the residual imprint is possible.
It was demonstrated that the increase in dissolution rate is intimately linked to the underlying structural changes induced by densification \citep{Hehlen2010, Deschamps2011}.

The aim of this paper is to provide new insight on the dissipative mechanisms at stake during the indentation process on a silica glass, using a very much enhanced version of the chemical dissolution technique introduced by \citet{Niu2012} introducing: (i) a chemical reactivity model based on the scientific literature and described to propose possible densification profiles underneath the apex of a residual indent, and (ii) the principle of geometrical similarity in sharp indentation  used to drastically increase the amount of information.
This method allows for the comparison of the chemical probe results with densification profiles published in the scientific literature, either from experiments or numerical simulations.

The paper is organized as follows.
We will first show that different constitutive equations along with FEM simulations are able to macroscopically agree with indentation experiments while describing different behaviours beneath the imprint.
Then, we describe the chemical dissolution technique and the way the principle of geometrical similarity is used in indentation testing.
Finally, the results of the chemical dissolution technique are compared to results either issued from numerical simulations or from micro Raman spectroscopy mappings.
The differences are discussed at the light of Raman resolution and possible convolution artefacts.

\section{Numerical simulations of the indentation test}
\label{sec:fea}

\begin{figure}[ht]
    \centering
    \includegraphics[width=.99\textwidth]{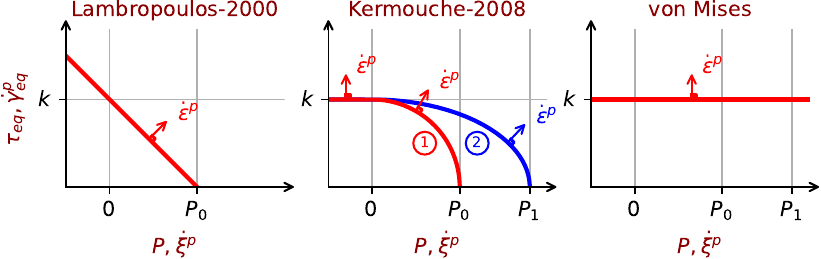}
    \caption{Description of the different constitutive models in the shear ($\tau_{\textsf{eq}}$) - pressure (P) plane.
    The red line corresponds to the initial yield surface while the blue one corresponds to hardening.
    The direction of the plastic strain rate,
    is also superimposed in the isochoric plastic strain rate ($\dot{\gamma}^p_{\textsf{eq}}$) vs densification rate ($\dot{\xi}^p$) plane and represented by arrows (see text for details).
    The models of Bruns et al. ressemble that of Kermouche et al..}
    \label{fig:models}
\end{figure}

In this section, we select four different constitutive equations that have been proposed in the literature to describe the mechanical response of silica glass to indentation and perform a numerical simulation of this test.
All models (we referRefs. \citep{Xin2000, Kermouche2008, Bruns2017, Bruns2020} for the details of the models as well as the values of the material parameters) are rate-independent and use a yield criterion and a flow rule.
The former is plotted in the equivalent shear (\(\tau_{\textsf{eq}}=\sqrt{\frac{1}{2}\, \text{tr}\, (\ten{s}\cdot\ten{s})}\), \(\ten{s}\) is the
deviatoric part of the Cauchy stress tensor) - pressure (p) plane.
The latter (the direction of the plastic strain rate \(\ten{\dot{\epsilon}}^p\)) is superimposed in the equivalent isochoric shear plastic strain
rate (\(\dot{\gamma}^p_{\textsf{eq}} = \sqrt{2\, \text{tr}\, (\dot{\ten{\gamma}^p}\cdot \dot{\ten{\gamma}^p})}\), $\dot{\ten{\gamma}^p}$ being the plastic shear strain rate) vs densification rate (\(\dot{\xi}^p = -\text{tr}\, \ten{\dot{\epsilon}}^p\), tr being the trace operator) plane.
Details of the constitutive equations and numerical procedures can be found in Supplemental A, but their salient mechanisms are described in \cref{fig:models}.

\begin{figure}[ht]
    \centering
    \includegraphics[width=\textwidth]{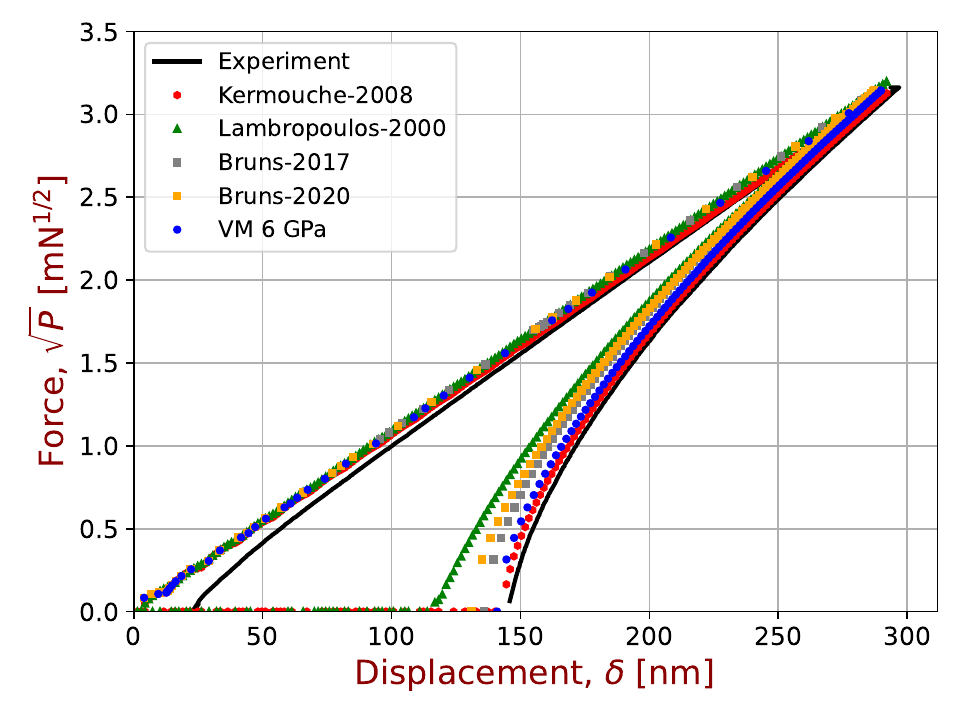}
    \caption{Comparison of the mechanical response of the indentation test on SiO$_2$ between experiment and three selected constitutive models.
        The experimental curve has been shifted to the right by 25 nm to account for the truncated indenter.
        The number of markers for the plots has been reduced for clarity.
        The choice of plotting the square of the force rather than the force itself follows the same idea.}
    \label{fig:Ph}
\end{figure}

For all the models, the force-displacement curves $\left(P,\delta\right)$ are shown in \cref{fig:Ph} along with experimental results presented later in \cref{sec:mat}.
The latter have been shifted to the right by $\Delta \delta = \unit{20}\nano\meter$, the truncated length, to account for the blunted indenter section (see \cref{sec:mat}).
In other words, it gives the mechanical response of the experimental data had the indenter been perfect (for depths greater than  2 to 3 times $\Delta \delta$).
Overall, the numerical results of all models show very close agreement with the experimental data.
The VM model (or J$_2$-plasticity, with a yield strength of $\unit{6}\giga\pascal$) and the Kermouche model show the best comparisons.
The model of Lambropoulos \etal~ \citep{Xin2000} does not match exactly the experimental curve.
This is in contrast with the results from Ref. \citep{Xin2000} and is explained by the $\unit{20}\nano\meter$ shift of the data and the low penetration depth.
For the model of Kermouche \etal\ \citep{Kermouche2008}, there is a close match with the experimental data and it does not suffer from the $\unit{20}\nano\meter$ shift since, in their work, high penetration depths $\delta_m = \unit{2}\micro\meter$ were used for parameter identification so that this shift does not play a crucial role (for the first model, $\delta_m = \unit{500}\nano\meter$).
The reasons for the small discrepancy between simulation and experimental data for the model of Bruns are the same as those for that of Lambropoulos (see \citep{Bruns2020}).

\begin{figure}[ht]
    \centering
    \includegraphics[width=.8\textwidth]{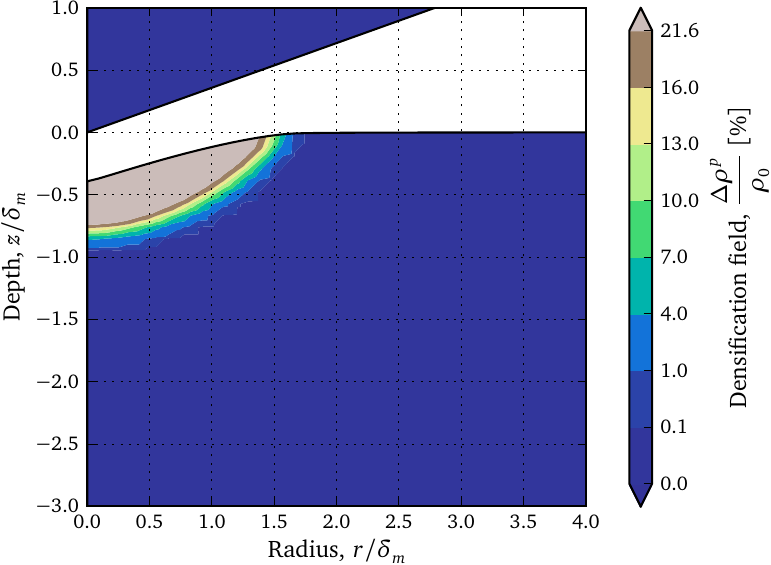}\\
    \includegraphics[width=.8\textwidth]{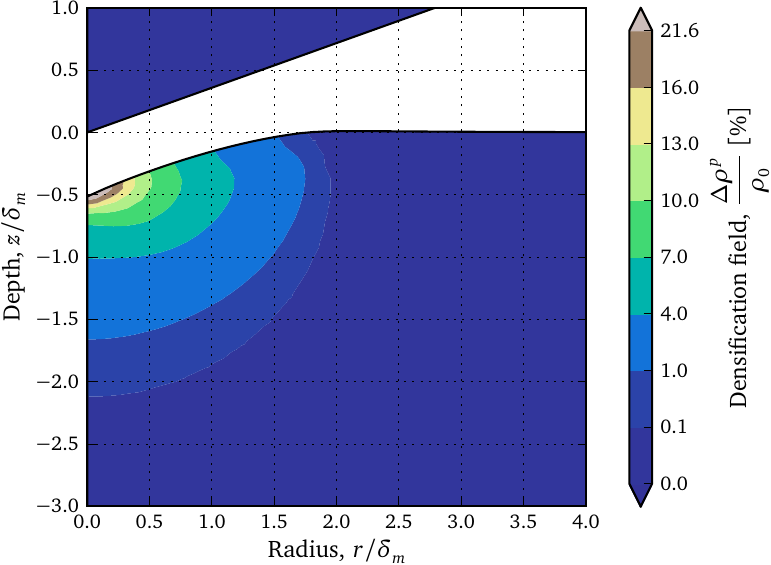}
    \caption{Densification field underneath the imprint with the model of Lambropoulos \etal. (top)
        and  Kermouche \etal. (bottom)
        Dark blue zones are not densified and light brown ones are above the experimental saturation value of 21.6\%.
        The scale is non linear to highlight the isolines at 1 \permil\ and 1 \%.
        The densification field ($\tau_D$) is not homogeneous in the zone above 21.6\% and reach unphysical values of more than 100\%.}
    \label{fig:Prints}
\end{figure}

The densification field underneath the indentation imprint, after unloading, is shown for the models of Lambropoulos and Kermouche in \cref{fig:Prints}.
The model of Bruns \etal~is similar to that of Kermouche and of course the VM model does not exhibit any densification.
The densification levels are presented with a non linear scale from 0 (no densification) to 21.6\% (saturation in densification).
Two colours (blue and light brown) indicate no densification or values above the saturation level, respectively.
The latter can be due to the fact that the models do not account for saturation as seen in \Cref{fig:Prints}.
In the following, we take the iso-value of $0.1 \%$ in densification as the boundary of the densification zone.
The model of Lambropoulos \etal\ describes a densification zone that extends up to $0.9 \delta_m$, with a core for which densification values are higher than $21.6\%$ which  is not homogeneous since the model does not account for saturation in densification.
The model of Kermouche \etal~describes a densification zone that extends up to $2.1 \delta_m$.
Inside this zone the densification is not homogeneous and there is a smooth gradient from underneath the indenter tip to the non-densified zone.
While this model does not account for the saturation in densification, it has a negligible impact on the densification fields.

Different constitutive equations are therefore able to reproduce the macroscopic mechanical response observed in indentation. 
A model that does not take densification into account would actually be able to do this. 
On the other hand, the densification field obtained below the surface of the residual imprint is very strongly impacted by the description of densification in the constitutive equations.
Therefore, it is necessary to bring new experimental data on what takes place beneath the indenter tip in the material, to discriminate among different models.
This is the objective of the next Section.

\section{Experimental}
\label{sec:experimental}

\subsection{Material and instrumented indentation testing}
\label{sec:mat}

A commercial silica glass (SiO$_2$ 99.6 mol $\%$, Spectrosil\texttrademark ) from Saint Gobain company (France) was used in this study.
The glass surface was polished with cerium oxide, and subsequently annealed for $\unit{2}\hour$ at the glass transition temperature ($T_g$ = 1100°C).
The mass density of the glass after annealing was $\rho_0 = \unit{2.2}\gram\per\centi\meter\cubed$.
One pristine (indentation free) sample of the glass was retained for the dissolution rate measurements in its relaxed state.

Instrumented indentation tests were carried out with a nano-indenter testing device (TI950, Bruker, USA) at ambient conditions (23°C and 55\%relative humidity).
The indenter tip is a modified Berkovich diamond pyramid.
Both AFM imaging and a standard indenter tip calibration method on a fused quartz standard sample \citep{Oliver2004} lead to an indenter tip radius value of
about 260 nm.
Another way to qualify the bluntness of the tip is to calculate a truncated length \citep{Loubet1984,Keryvin2017n}.
The mechanical response of the indentation test is the force $P$ vs.~the displacement $\delta$ (counted positively).
The truncated tip defect length, $\Delta \delta$, is obtained by plotting $\sqrt{P}$ vs. $\delta$ for the fused quartz reference sample during the loading stage (increasing $P$).
This curve should be linear with its origin at (0,0) for a perfect tip (self similarity of sharp indentation, see \cref{sec:similitude}).
This is not the case for shallow depths below $\unit{50}\nano\meter$, so $\Delta \delta$ was calculated by taking the intercept of a linear fit of this curve for high values of $\delta$, as seen in \cref{fig:disp}.
Finally, $\Delta \delta$ is found to be $\unit{20}\nano\meter$.

\begin {figure}[ht]
\centering
\includegraphics[width=\textwidth]{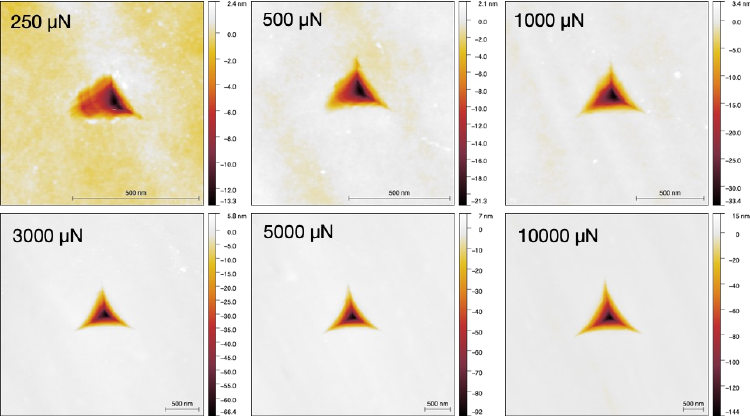}
\caption{AFM images of representative Berkovich residual imprints left on a polished then annealed silica glass surface.}
\label{fig:afm}
\end{figure}

Nano-indentation tests were carried out on a dedicated sample, with a `10-10-10' loading sequence: 10 s to reach the maximum load $P_m$, $\unit{10}\second$ of holding time, and $\unit{10}\second$ to unload the sample's surface.
They were load-controlled and the $P_m$ values ranged from $\unit{250}\micro\newton$ to $\unit{10}\milli\newton$.
The maximum and residual displacements are referred to as $\delta_m$ and $\delta_f$, respectively the former being measured by the nano-indenter system while the later is measured by atomic force microscopy (AFM).
Due to the high reproducibility of the nano-indentation test on the glass surface, five indents per chosen maximum load were performed.
All imprints, as imaged by AFM, were free of corner cracks, \textcolor{red}{as seen in \Cref{fig:afm}.}

\subsection{Chemical dissolution technique}
\label{sec:chemprob}

\begin {figure}[ht]
\centering
\includegraphics[width=\textwidth]{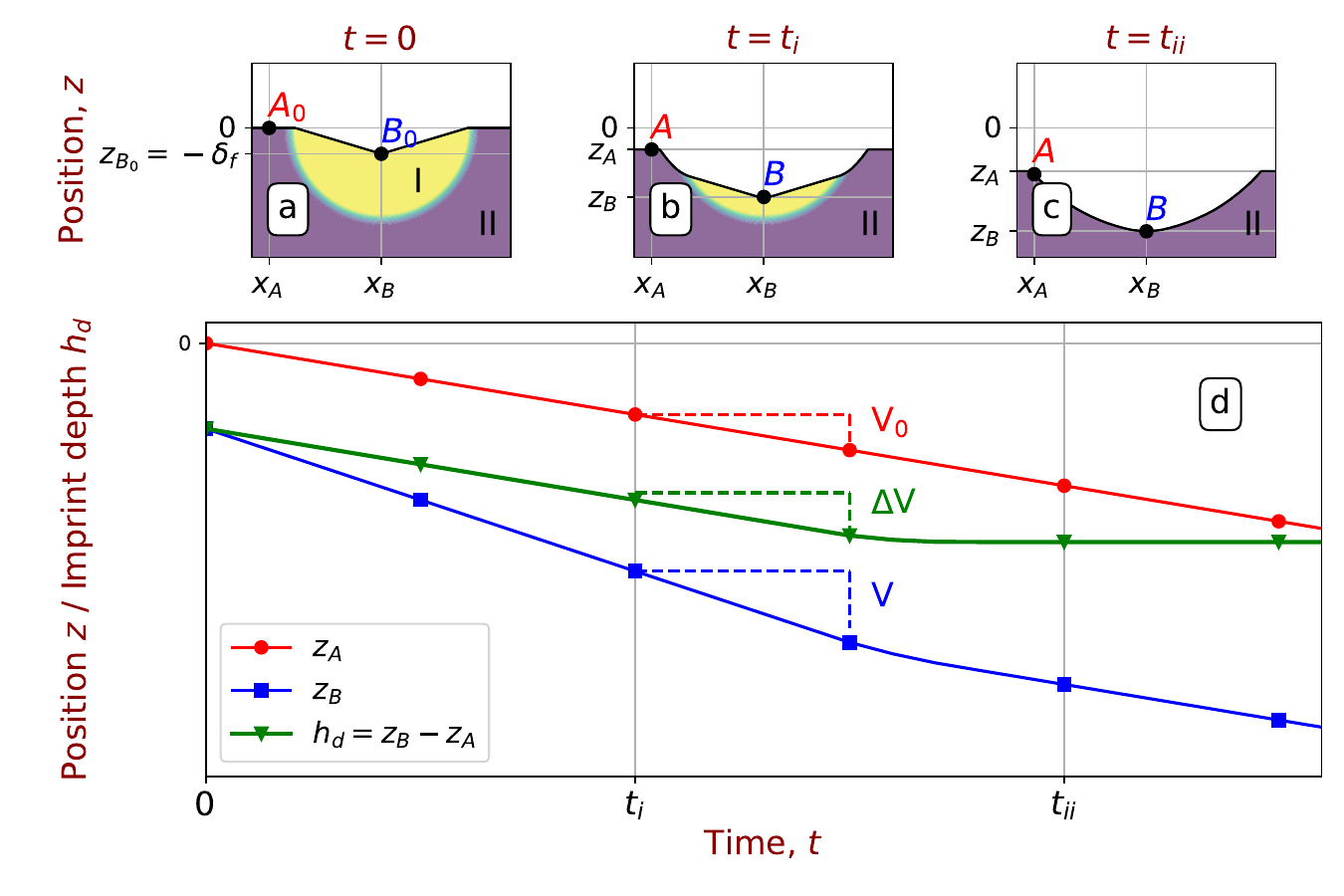}
\caption{Schematic of the dissolution process. The pristine sample exhibits a flat surface. \textcolor{red}{Colors red and blue are associated to points A and B, respectively}. \textbf{(a)} A sharp indentation test is performed on the surface leaving a residual imprint with a depth $\delta_f$. The dissolution rate is locally increased in the densified zone. At this moment noted $t=0$, the dissolution process is started. \textbf{(b)} The dissolution rate being higher at the bottom of the imprint, its apparent depth $h_d = z_{B} - z_{A}$ is increased during this first stage (this situation corresponds to dissolution steps till the bottom of the densified zone is reached). \textbf{(c)} When the dissolution of the densification zone is completed, the dissolution rate becomes homogeneous again and the apparent depth of the imprint $h_d = z_{B} - z_{A}$ stops changing. \textbf{(d)} Changes with time of $z_A$, $z_B$, $h_d$ (measured by AFM) for situations (a) to (c) \textcolor{red}{and associated slopes (dissolution rate of pristine glass, point A, V$_0$, that of point B, V, and their difference $\Delta$V=V-V$_0$).}}
\label{fig:schema}
\end{figure}

The indented glass sample was immersed in a Teflon\textsuperscript{TM} container filled with $\unit{50}\milli\liter$ of a 0.1 M NaOH solution heated to 80°C.
The temperature was kept at $\pm 0.5$°C in a thermally regulated furnace.
The prepared solution was divided into two separate containers for the indented and indent-free samples, respectively.
The latter was used to compute the dissolution rate $V_0$ by the weight loss method.
Those conditions allowed one to avoid saturation conditions throughout dissolution, as confirmed from the linear trend of the weight mass loss of the sample versus dissolution time \citep{Niu2012}.
The indented samples immersed in the alkaline solution were taken out periodically (every half to one hour) and rinsed consecutively with deionised water and ethanol prior to carrying out AFM measurements, which made it possible to record the three-dimensional geometry of the imprint after each dissolution stage.
Images were captured with the tapping mode of the AFM (Bruker, Nanoscope V, USA) equipped with silicon tips (TAP 300 Al) which apical angle is 70°C and the tip radii are no larger than \unit{10}\nano\meter.
Due to both a smaller tip radius and a sharper apical angle, when compared to that of the indenter, the geometry of the imprint was not altered by the finite size and geometry of the AFM probe tip.
Prior to carrying out any measurements, the AFM was calibrated with several grids: a 10 µm pitch of 200 nm deep squared holes and a \unit{3}\micro\meter\ pitch of $\unit{23}\pm\unit{1}\nano\meter$   deep engraved features (TGZ1).
Moreover, to limit the effect of typical AFM artefacts on the measurements such as thermal drift and piezo creep, thermal equilibrium of the system was established (about \unit{2}\hour) before capturing an image.
The size of the scanned area was large enough so that a sufficient area unaffected by the indentation process exists and may be used as a reference surface (\emph{i.e.}, set to zero tilt and zero offset).
For each loading condition, the topography of three to five indentation imprints were recorded by AFM as a function of dissolution time.
Following the technique described in Ref. \citep{Niu2012}, we focus on the evolution of the depth of the residual indentation imprint as a function of dissolution time.
For a clearer understanding of the process, a schematic is given in Figure \ref{fig:schema}.
The sample is shown in Fig. \ref{fig:schema} (a) after indentation with a residual imprint of depth equal to $\delta_f$.
Two points of interest are labelled $A_0$ and $B_0$, respectively, and their relative height is monitored by AFM $z_{A_0} - z_{B_0} = \delta_f$.
As shown in \cref{fig:schema} (b,c), throughout the dissolution process, $A_0$ and $B_0$ move to $A$ ($z_{A}$) and B ($z_{B}$), respectively, with the indentation imprint depth, $h_d$ at dissolution time $t$ being expressed as (and further referred to as the imprint depth):

\begin{equation}
    h_{d}(t) = z_{B}(t) - z_{A}(t)
    \label{eq:depth}
\end{equation}

Thus, at time $t=\unit{0}\second$ (no dissolution), $h_d(0)=-\delta_f$ and for dissolution times $t>0$, $|h_{d}(t)|>\delta_f$ because of the enhanced dissolution rate of the densified zone (coloured zone) that was generated during the indentation process.
Since the dissolution rate is constant over time for A, its position, from the initial point $A_0$ at the free surface ($z=0$), at dissolution time $t$, is known \emph{via} $z_{A}(t) = V_0\times t$ ($V_0 < 0$).
Therefore, the position of point B with respect to that of the initial free surface $z=0$ is expressed as:

\begin{equation}
    p\,(t) = V_0\, t + h_{d}(t)
    \label{eq:diss}
\end{equation}

The first term is evaluated by determining $V_0$ with the loss weight method on the pristine (indent-free) sample and the second term is measured by
AFM \emph{via} Eq. \eqref{eq:depth}.

The weight loss method on the pristine glass sample (indent-free) gives a dissolution rate of $V_0 = -71.7 \pm \unit{2.5}\nano\meter\per\hour$.
\textcolor{red}{The dissolution rate is computed from the mass loss of a pristine silica sample in the exact same solution as the one used for the chemical probe study and under the exact same temperature conditions. Several measurements are made over different time steps, a linear regression analysis is performed to compute de dissolution rate, the uncertainty comes from the error analysis associated with the linear regression analysis.}
\Cref{fig:dissolution} (a) reports, for indentation maximum loads $P_m$ ranging from $\unit{0.25}\milli\newton$ to $\unit{10}\milli\newton$, the evolution of imprint depths $h_d$ as a function of dissolution time $t$.
Regardless of $P_m$, the curves exhibit similar behaviour.
In the first zone, referred hereafter as Region I, the depth of the imprint (its absolute value) increases with dissolution time.
It means that, in this region, the dissolution is faster than that of the free surface far from the imprint.
For longer dissolution times, a second zone referred hereafter as Region II, is characterized by a plateau, for which the imprint depth remains constant.

\subsection{Principle of geometrical similarity}
\label{sec:similitude}
We use in what follows the principle of geometrical similarity for sharp indenters.
Details can be found in Supplemental B.
As a simple example at $t=0$ the similarity of the indentation test for two maximum forces ($P_m$) and residual depths ($\delta_f$) implies, considering two indentations 1 and 2 $(P_{m,1}<P_{m,2})$:

\begin{equation}
    \dfrac{ \delta_{f,1} }{ \sqrt{P_{m,1}} }=\dfrac{\delta_{f,2}}{\sqrt{P_{m,2}}}=\delta_f^*=\text{constant}
    \label{eq:dim_an_2}
\end{equation}

\noindent where the notation \textsf{property}$^*$ will be used in the rest of the manuscript to express this property within the similarity framework (i.e.~divided by its respective $\sqrt{P_m}$). 
We found, for the experiments shown in \cref{sec:mat}, $\delta_f^*$ = 1.51 $\pm$ 0.03 nm/$\mu$N$^{1/2}$ and $\delta_m^*$ = 2.86 $\pm$ 0.02 nm/$\mu$N$^{1/2}$.
Regarding the chemical probe (CP) data, since dissolution rates values are invariant through similarity, the time $t_1$ required to dissolve the densified volume of indentation 1 will be smaller than $t_2$ and will satisfy the equation:

\begin{equation}
    \dfrac{t_2}{\sqrt{P_{m,2}}} = \dfrac{t_1}{\sqrt{P_{m,1}}}=t^*=\text{constant}
    \label{eq:dim_an_3}
\end{equation}

\noindent As a consequence the imprint depth (\cref{eq:depth}) through dissolution satisfies:

\begin{equation}
    \dfrac{h_{d,2}(t_2)}{\sqrt{P_{m,2}}}=\dfrac{h_{d,1}(t_1)}{\sqrt{P_{m,1}}}=h_d^*=\text{constant}
    \label{eq:dim_an_4}
\end{equation}

\begin {figure}[ht]
\centering
\includegraphics[width=\textwidth]{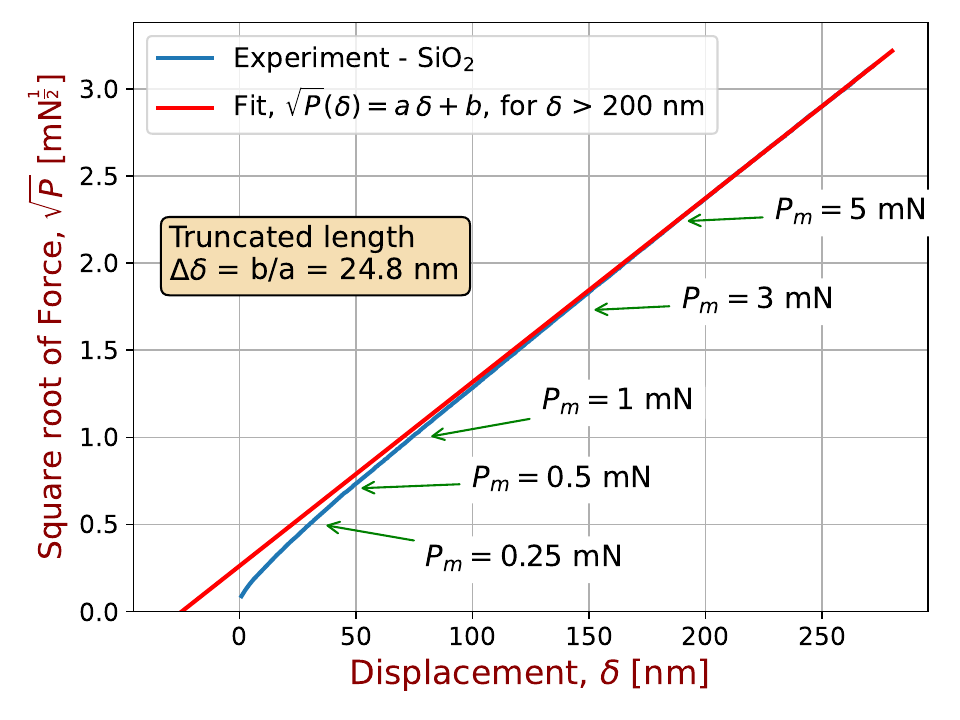}
\caption{Evolution of the square root of the force versus the indentation depth during the loading stage of a 10 mN indentation test on the silica glass.
    A linear fit for depths higher than 200 nm (for which we are in the similitude regime) is extrapolated down to the x-axis to give the tip defect in terms of a truncated length $\Delta\delta $ $\sim$ 25 nm. The positions of the lower loads tested are also indicated for discussing the effect of the imperfect tip.}
\label{fig:disp}
\end{figure}

Thus, one can transpose easily the experimental data (space and time) into the similar framework and, if needed, rescale the data set to any load by simply dividing both space and time by the square root of the considered load (\cref{eq:dim_an_3,eq:dim_an_4}).
The inverse allows also to re-scale the so obtained master curve  to any desired indentation load.
It makes it possible to extract complementary data from multiple indentation-dissolution tests and increase drastically the precision of the method.
Furthermore, to circumvent the difficulty for correcting the impact, at low loads, of the truncated length (see \cref{fig:disp}) on residual imprint depth measurements, the choice is made to consider the \emph{cumulative increase of the imprint depth} as a function of time ($\Delta h_d$) defined by :

\begin{equation}
    \Delta h_d(t) = h_d(t) - \delta_f
    \label{eq:dim_an_6}
\end{equation}

Within the GS framework it becomes :

\begin{equation}
    \Delta h_d^*(t)=\dfrac{ h_d(t) - \delta_f}{\sqrt{P_m}}
    \label{eq:dim_an_7}
\end{equation}

\begin {figure}[ht]
\centering
\includegraphics[width=\textwidth]{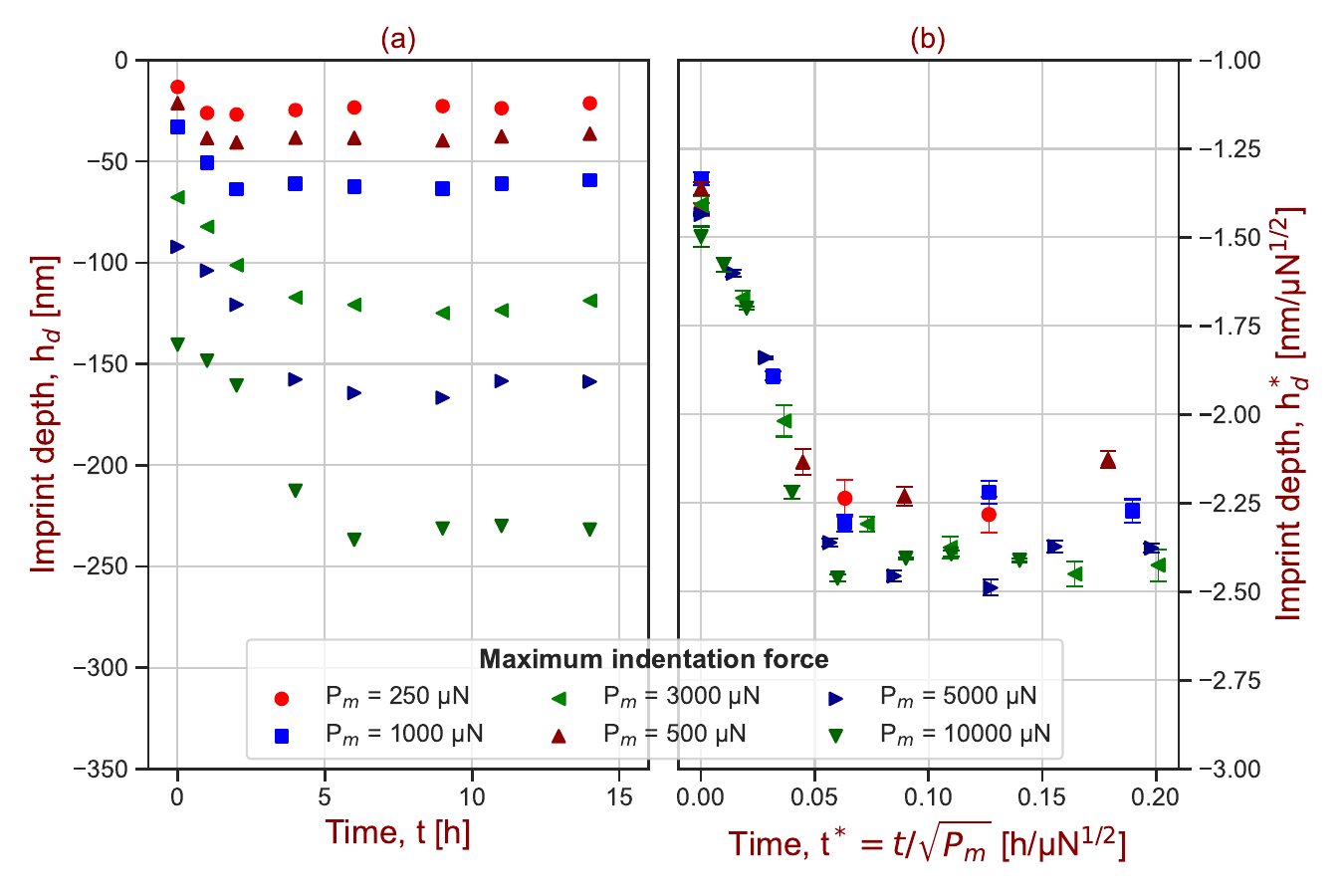}
\caption{Evolution of the residual imprint depth (h) as a function of dissolution time (t) for Berkovich indentation loads ranging from 250 $\mu N$ to 10 $mN$ on silica glass. a) as measured by AFM; b) same data but rescaled using the geometrical similarity of the indentation test (i.e. both distance and time are divided by the square root of the considered maximum force). The error bars are plotted only for (b) for sake of clarity.}
\label{fig:dissolution}
\end{figure}

We therefore divide all the data (both depth and time) of Fig. \ref{fig:dissolution} a), according to \cref{eq:dim_an_3,eq:dim_an_4}, by the square root of the maximum indentation applied force $\sqrt{P_m}$.
It is clear from \cref{fig:dissolution} b) that, in most cases, the data points collapse into a single master curve presenting the two regions defined previously.
However, there are two notable exceptions at low loads of \unit{0.25}\milli\newton\ and \unit{0.5}\milli\newton.
This is not surprising since, as highlighted in \Cref{fig:disp}, for low loads (and hence low displacements), we are not in the similitude regime anymore because of the roundness and the truncated length of the pyramid.

\begin {figure}[ht]
\centering
\includegraphics[width=\textwidth]{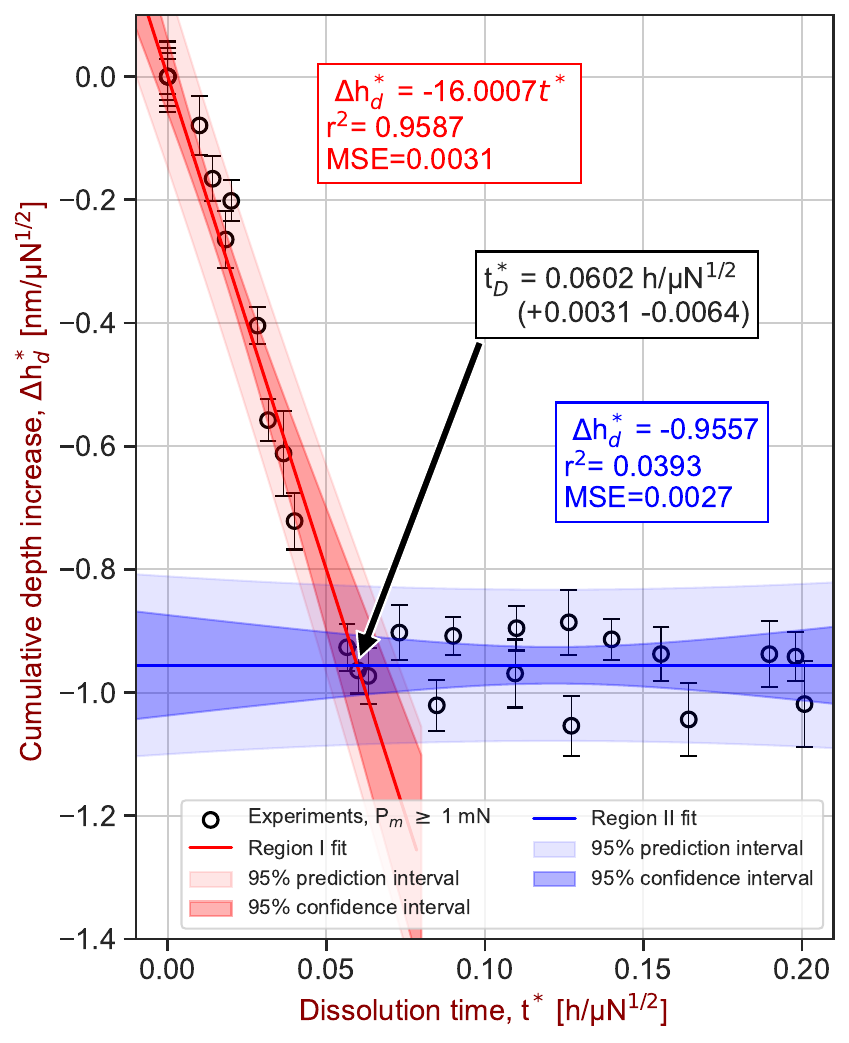}
\caption{Cumulative depth increase $\Delta$h$_d^*$ of the residual imprint through dissolution time for Berkovich indentation loads on silica glass  ranging \textcolor{red}{from 1 mN to 10 mN} reported within the geometrical similarity framework of the indentation test.
    The confidence band is the confidence region for the correlation equation and the prediction band is the region that contains roughly 95\% of the measurements.
    r$^2$ is the correlation coefficient  and MSE is the mean squared error (the smaller the better the model fits the data).}
\label{fig:cum_depth}
\end{figure}

We have removed these two data sets from our further analysis and plotted $\Delta h_d^*$ with respect to $t^*$ in \cref{fig:cum_depth}.
Region I was well fitted by a least squares linear regression.
This indicates that, in this region, the increase in dissolution rate is rather homogeneous and is found to be $\Delta\,V$ = 16.0 $\pm \unit{1.1}\nano\meter\per\hour$.
Region II was fitted by a plateau value of $\Delta h^*_D$ = -0.96$\pm$ 0.04 $nm/ \mu N^{1/2}$.
These fits are superimposed to the experimental data in \cref{fig:cum_depth} together with 95\% confidence intervals.
The confidence band is the confidence region for the correlation equation and the prediction band is the region that contains roughly 95\% of the measurements \cite{Altman2015}.

These extended results makes us assume that the densification zone (Region I) is rather homogeneous followed by a steep transition to the non-densified zone (see also the discussion part on the reactivity model here after).

Hence, knowing the dissolution rates of the free surface (point $A$ in Fig. \ref{fig:schema}) and the one of the densified region (slope of region I) we can estimate the thickness of the densified glass under the residual imprint apex that was dissolved at time $t_D^*$ (i.e.~the position of the boundary of the densified zone beneath the apex of the residual imprint)
to be $\delta^*_D$ =-5.28 $\pm$ 0.53 nm/$\mu$N$^{1/2}$ or with respect to $\delta_m^*$ (respectively $\delta_f^*$) it gives 1.85 ( respectively 3.50).

\section{Discussion}\label{sec:discussion}

\subsection{Comparison to Raman mappings}
\label{sec:raman}

\begin {figure}[ht]
\centering
\includegraphics[width=\textwidth]{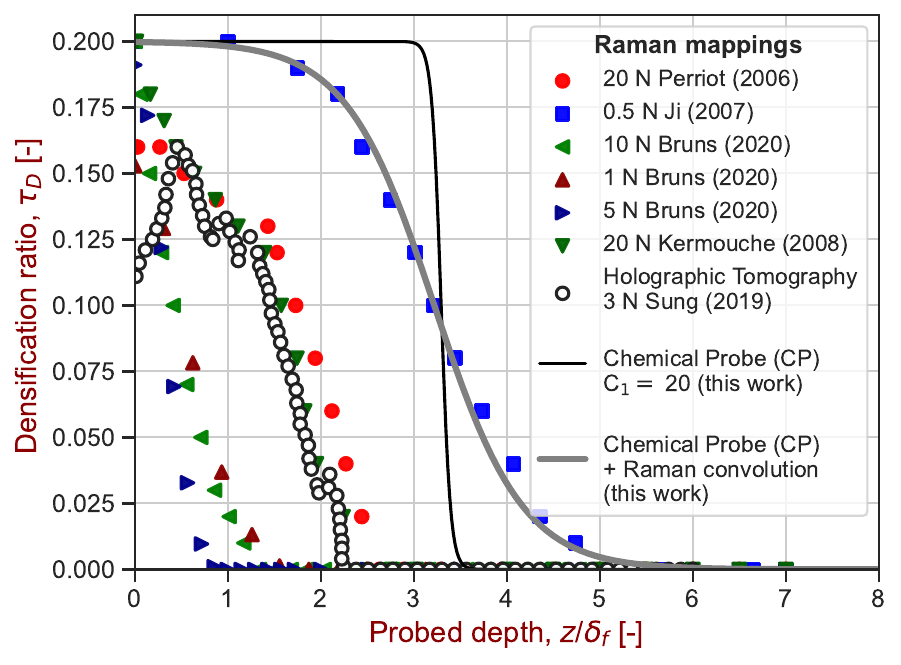}
\caption{Raman profiles from the literature of densification profiles versus probed depth underneath the residual Vickers indentation imprint apex normalized to the residual depth of the imprint $\delta_f$ ($z=0$ at the tip of the imprint).
}
\label{fig:profiles_lit}
\end{figure}

The selected constitutive models are shown to reproduce accurately the mechanical response of the indentation test, \emph{i.e.} the force--displacement curve, although there are significant differences between models.
In other words, matching the experimental $P$-$\delta$ is necessary but not sufficient.
This known fact has led researchers to compare their simulation results to Raman micro-spectroscopy mappings of the densified domain at the residual indentation imprint.
Nonetheless the size of the Raman probe usually implies the use of much higher indentation loads ($\unit{500}\milli\newton$ to $\unit{20}\newton$ in the literature) when compared to this study.
From Refs. \citep{Perriot2006, Ji2007, Kermouche2008, Bruns2020},  four such densification profiles below a residual Vickers indentation imprint may be extracted and plotted on the same graph in \Cref{fig:profiles_lit} as a function of $\textstyle \frac{z}{\delta_f}$, the probed depth beneath the imprint apex ($z$), normalised to the residual imprint depth ($\delta_f$), using the similitude principle\footnote{Ref. \citep{Kermouche2008} being a new post treatment of Ref. \citep{Perriot2006}}.
A fifth profile obtained from a digital holography tomography (DHT) method \cite{SATOSHI2019} is also added.
The most striking observation is that a large disparity in terms of densification landscapes does exist although it is the same material and the same indenter shape.
The only consensus lies in the maximal densification ratio reached right underneath the residual imprint apex ($\sim$ 20 \%).
\citet{Bruns2020} report for loads equal to 1, 5, 10 N a rather shallow densified region ($z/\delta_f$ $\approx$ 1) with a quite steep decrease.
\citet{Kermouche2008} show also for a 20 N load a continuous decrease of the densification ratio, yet over a larger depth ($z/ \delta_f$ $\approx$ 2.5).
Despite a lower densification ratio at shallow depth $(z/ \delta_f<0.5)$ the DHT profile from a 3 N indentation laod is very similar to the 20 N one \cite{Kermouche2008}.
Finally, \citet{Ji2007}, for a 500 mN load, exhibits a constant and maximal densification ratio over a certain depth before a continuous decrease of the latter before retrieving the pristine glass at an even larger depth ($z/\delta_f$=5).

\subsection{A Densification-Reactivity Model: origins of the dissolution rate increase}
\label{sec:reactivitymodel}

Because of these discrepancies and the difficulty to straightforwardly compare those profiles to the chemical probe experimental data, it is of great importance to investigate the possible origins of the dissolution rate increase in the densified zone and try to gather more information out of it.
The variation of the dissolution rate ($\Delta V$) in the densified silica glass area, which is experimentally observed is the resultant of two components: a geometrical one and a structural one.\footnote{Please note that despite these factors have an impact on the dissolution rate, the effect of: residual stresses (discussed in \citep{Niu2012}); local surface curvature; mass transport from the surface to the bulk solution, are neglected in the following discussion as their effect is at worst one order of magnitude lower than the considered variations.}

\begin{equation}
    \Delta V=\DVG+\DVR
    \label{eq:DV}
\end{equation}

The geometrical component $\DVG$ finds its origin in the fact that more atoms are packed within the same volume at iso Si-O bond number, at least for silica glass.
Hence for the same volume more bonds will have to be hydrolyzed in the densified glass for a complete dissolution.
Thus its contribution to the dissolution rate will be negative (i.e.~a decrease),

\begin{equation}
    \frac{\DVG}{V_0} (\tau_D) = \frac{- \tau_D}{1+\tau_D}
    \label{eq:DV3-bis}
\end{equation}

Details for obtaining this equation can be found in Supplemental C.
The structural (and positive) component ($\DVR$) finds its origin in the structural modifications that result from the densification process.
The latter is extensively described by \citet{Hehlen2010} and the full concept is explained in Supplemental C.
The following expression is found:

\begin{equation}
    \dfrac{\DVR}{V_0} (\tau_D) =  e^{\alpha \cdot \tau_D}-1
    \label{eq:DVreact_bis}
\end{equation}

where $\alpha =1.771$ is a calibration constant.

\textcolor{red}{
In a previous paper \cite{Niu2012}, Vickers indentation imprints were partially recovered using a specific thermal annealing procedure (0.9 Tg (K) for 2h), which allows for a complete densification recovery (the Raman signal is back to that of the pristine glass) but not all permanent strains (volume conservative ones also referred to as shear flow). 
In doing so the dissolution rate enhancement observed on post indentation imprints totally disappeared after such a thermal treatment. 
The dissolution rate enhancement is therefore strongly associated with the silica network changes resulting from the densification process.
Let us note also that silica is a “relatively simple” case; for other glasses such as soda-lime silica glass  \cite{Niu2012} or borosilicate glass other structural modifications impacting the dissolution rate are at stake (Qn species variation, oxydation number modification…).
}
\subsection{Geometrical Similarity, Raman profiles and comparaison to CP experimental results}
\label{geometrical-similarity-raman-profiles-and-comparaison-to-cp-experimental-results}

In order for Raman spectroscopy results to dialogue with the CP ones, $\Delta$h$_d^*$ versus $t^*$,
both the time scale (i.e.~the time needed to reach a considered depth) and the total imprint depth increase, which occurred during this time lap, need to be computed. This technical aspect is described in Supplemental D and in \Cref{fig:synopsis}.
This method is applicable to any densification profile (Raman, FEM, DHT).
Once it is done all the data may be plotted on the same   $\Delta$h$_d^*$ versus $t^*$ graph as it is illustrated in \Cref{fig:dh_all}.
The first observation is that all curves  present a similar trend (i.e. a decrease of  $\Delta$h$_d^*$ followed by a plateau) but differences do appear firstly in the final position of the plateau and secondly in the sharpness of the transition zone.
Only one densification profile \citep{Ji2007} from a Raman spectroscopy mapping of a 500 mN Vickers indentation is very close to the chemical probe results.
The other profiles fell short by more than 30\% for the closest ones \citep{Xin2000, Perriot2006,SATOSHI2019} regarding the final cumulative depth increase value and by 80\% for the furthest ones \citep{Bruns2017, Bruns2020}.
This difference finds an explanation both in the existence and in the size of a densified zone having a constant and maximal densification ratio of about 20\% right underneath the apex of the residual imprint.
Regarding this latter point, the CP technique is in agreement with Raman observations carried out on indentation imprints made in the 100 mN range \citep{Gerbig2020, Ji2007} but differs drastically from those made at much higher loads ($1$ to $\unit{20}\newton$).
Such a discrepancy may find its origin in two different aspects:

\paragraph{The load and similitude regime} Although we have shown that the similitude regime may be lost at very low loads it can be lost also above a certain load threshold at which other energy dissipative mechanisms such as cracks dissipate a non negligible amount of energy that is not dissipated by plasticity mechanisms anymore. A new length scale is therefore involved.

\paragraph{The confocal volume of Raman spectroscopy} From which the densification ratio is computed (and averaged) at a certain depth is a rather complex parameter to determine.
It depends on numerous yet rather different parameters such as the detection optics, the focusing optics (the effective pinhole), the laser wavelength, its energy profile. Moreover, it is also largely affected by any refractive index mismatch or variation of it as well as sample induced spherical aberration that may deteriorate the confocality of the system (both size and shape) \citep{Maruyama2010,Adar2010,Everall2010,GERBIG2024}.
Although this would need a specific study for more quantitative conclusions, the presence of opened cracks for instance in the densified zone may have a dramatic impact on the confocality of the system.
The work of \citet{GERBIG2024} provides quantitative results upon the bias encountered when probing the densified zone using Raman spectroscopy. The reader is referred to this very recent work for deeper insight regarding this topic.\\
\ \\

As a rule of thumb,  a finely tuned system (i.e.~with an optimized confocality) on the one side and as few cracks as possible (refractive index mismatch events) in the densification zone on the other side may ensure good enough conditions for accurate (yet representative of the GS regime) Raman densification ratio mapping at an indentation imprint site.

\subsection{A possible densification profile shape estimated from CP results?}
\label{sec:?}

This being said it is also of interest to identify a possible densification profile able to best describe the chemical probe experimental results for further discussion and comparison with the literature.

\begin {figure}[ht]
\centering
\includegraphics[width=\textwidth]{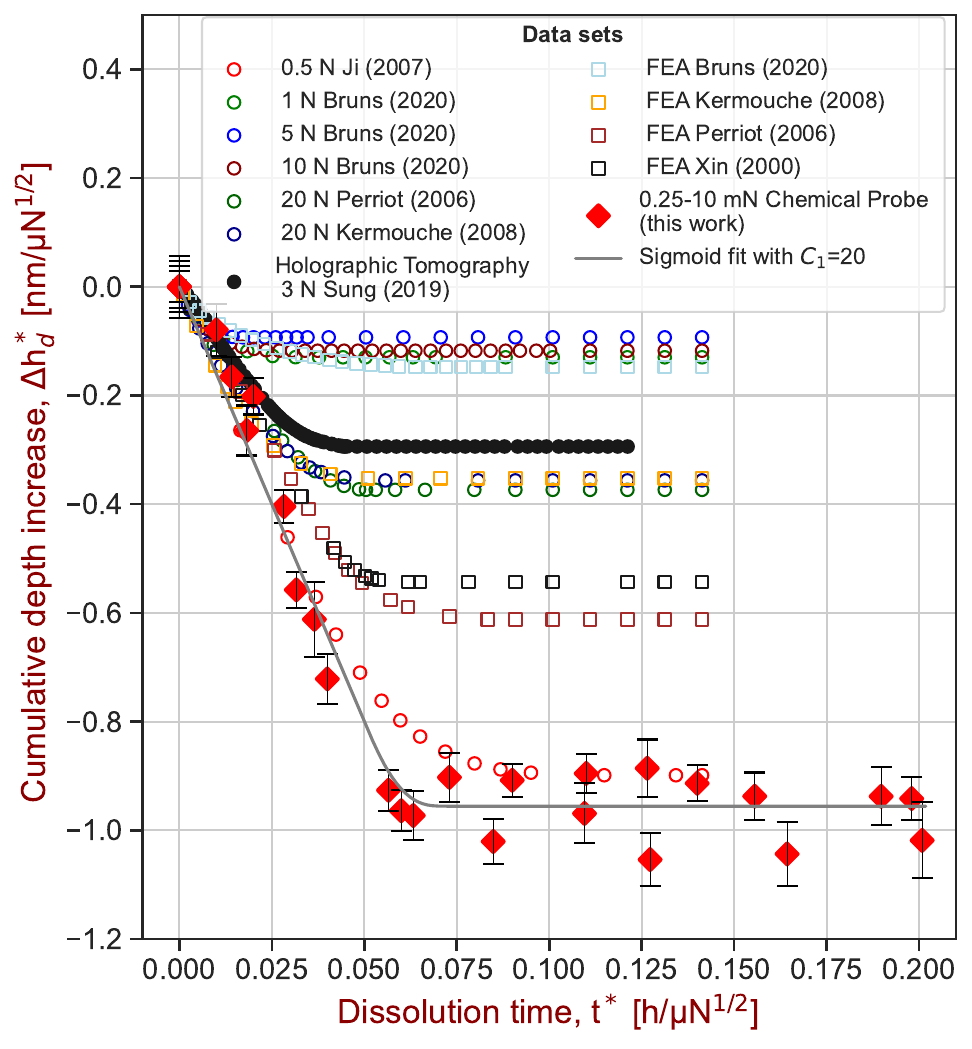}
\caption{Cumulative depth increase $\Delta$h$_d^*$ of residual indentation imprints versus time ($t^*=t/\sqrt{P_m}$) computed from  densification profiles available in the literature using the chemical reaction model developed in this work. Results are plotted along with experimental data from the chemical probe technique (opened red diamonds). The lines red and blue where computed from sigmoidal densification profiles so that : in red the profile was adjusted to best fit the CP experimental results; in blue  the profile was the best fit of the Raman densification profile. }
\label{fig:dh_all}
\end{figure}

The only possible solution that would satisfy a constant $\Delta V$ in region I is to have both $\DVG$ and $\DVR$ constant over this same region, which in turn implies the existence of a zone having a constant densification ratio underneath the indentation imprint
\footnote{This statement is in apparent disagreement with Raman studies (high loads) reported in \citep{Perriot2006, Kermouche2008, Bruns2020} as well as corresponding FEA studies reported in \citep{Perriot2006, Kermouche2008, Bruns2020}, but not with studies (low loads) reported in \citep{Ji2007, Gerbig2020}}.
This constant $\Delta V$ region is also logically followed by a transition zone defined by a negative densification gradient until the pristine glass that was not affected by the indentation process is retrieved.
We found that a  two parameters (C$_1$, C$_2$) sigmoidal function (\cref{eq:tau_D_sig}) with  $C_2=3.29$  the position of the transition and  $C_1\geq 20$  the width of the transition zone (see \cref{fig:dh_all}) is an excellent candidate that describes well the chemical probe data set  (cf.~\cref{fig:dh_all}).

\begin{equation}
    \tau_D\,(\frac{z}{\delta_f})=\tau_\textsf{max}\times \left( 1-\frac{1}{e^{-C_1 \cdot \left(\frac{z}{\delta_f}-C_2\right)}} \right )
    \label{eq:tau_D_sig}
\end{equation}

In this equation C$_1$ is the sole adjusting parameter for which the quality of the dataset does not allow for further refinements and C$_2$ is computed from experimental results so that the time  to reach the transition time $t^*_D$ computed from this profile equals the experimental CP one. ($\tau_\textsf{max}$ is the maximum densification ratio $\sim$ 20\%.)

From there it becomes possible to investigate upon the confocality of the system used for Ji's \citep{Ji2007} study as a possible explanation for the differences observed between CP and Ji in  \Cref{fig:dh_all} .
To do so we roughly estimated the z confocality to exhibit a Gaussian shape characterized by $\sigma$ the standard deviation that needs to be determined.
In this case the z resolution of the confocal system lies roughly between $2 \sigma$ and $4 \sigma$.
To determine $\sigma$, the product of convolution of the CP densification profile by the Gaussian function describing the confocality is computed and $\sigma$ is adjusted to best match the Raman densification profile.
It is found that the CP densification profile convolved by a Gaussian function having $\sigma$ =0.85 nm/$\mu$N$^{1/2}$ describes unexpectedly well the Raman densification profile.
After rescaling to a $\unit{500}\milli\newton$ indentation load to match the Raman experimental conditions the z resolution or confocality can therefore be estimated and it is found to lie between $\unit{1.7}\micro\meter$ (2$\sigma$) and $\unit{3.4}\micro\meter$ (4$\sigma$), which falls within expected experimental values.

\section{Concluding remarks}

A chemical dissolution technique was employed to investigate the densification process underneath an indentation imprint in silica glass.
It relies on a higher dissolution rate for densified areas w.r.t. non densified ones.
This technique allows one to extract high spatial resolution information (nm range), even for low-loads indentation tests carried out to avoid the onset of
spurious cracking events, in contrast to micro-spectroscopy techniques including Raman and Brillouin.
Taking into account the self similarity of the sharp indentation process made it possible to considerably extend the soundness of this method.

Applying this same post treatment using the similitude regime to data (i.e.~densification profiles) from the literature allows for the direct comparison of different densification mappings by Raman spectroscopy or digital holography tomography and highlighted a wide variability between studies.
Selected constitutive equations for silica glass from the literature have been used and the results of numerical simulations of the indentation process show also large differences in terms of densification profiles.

Using the extended version of the chemical dissolution technique reveals a densification zone very homogeneous (in terms of densification levels) along the vertical axis and extends up to $\sim$ 3 times the residual indentation depth with steep densification gradients between this zone and the non-densified one.

A reactivity model was then developed, that relates the dissolution rate increase to the densification ratio of the silica glass.
It makes it possible to describe, by a 2 parameters sigmoidal function, the densification depth profile.
Chemical probe results and those computed from Ji \citep{Ji2007} using Raman profile spectroscopy are found to be rather close.
Yet, the Raman densification profile, which is also adequately described by a sigmoidal function, exhibits a smoother transition.
Confocality or the vertical resolution of the Raman mapping provides an explanation for the differences between those two and convolving the densification profile computed from the chemical probe results by a Gaussian function results in a remarkable agreement between chemical probe and Raman mapping data.
From this work  it can be concluded that as long as both the similar behavior of the indented material and the geometrical similarity of the indenter are fulfilled, no effect of the load upon the experimental measurements, thus the densification profile, is evidenced, even for such small indentation depths. 

The present study fully agrees with a very recent study from \citet{GERBIG2024} and brings complementary proof regarding the effect of the Raman probe size on the so obtained densification profile.

\textcolor{red}{
The present work demonstrates that the only solution for measuring a constant dissolution rate over a large depth range in the plastic zone is that the densification ratio in this zone is constant and has reached the saturation level (i.e. 21\% densification ratio for silica glass also measured by Raman spectroscopy in several studies). The densification ratio between this extremum and that of the pristine glass might then be estimated for any dissolution rate increase with a reasonable margin of error having in mind both the progressive permanent structural modification with stress level and the sigmoidal shape of the densification ratio with stress (common to very different glass compositions). 
}

The geometrical self similar post treatment presented in this work may be used for other glass composition, yet the reactivity model will have to be adapted. It also opens up interesting possibilities for the full 3D investigation of the densification volume at nanoindentation site.

The results of this study are expected to provide a sound experimental database when trying to establish constitutive models or compare with other numerical methods used to simulate the indentation process in glass
\citep{Jebahi2013, Kilymis2013a, Yuan2014}.

\hypertarget{acknowledgements}{%
    \section*{Acknowledgements}\label{acknowledgements}}
\addcontentsline{toc}{section}{Acknowledgements}

VK would like to thank financial support from Univ. Bretagne Sud (CRCT) and Univ. of New South Wales for a sabbatical leave and that of French CNRS for a `delegation'.
JPG would like to thank the ANR for supporting a part of this work through the grant ANR-07-JCJC-037 as well as the Ministry of higher education for the PhD grant of KH.
CPER PRIN2TAN also contributed to fund this study.
We also would like to thank A. Moreac (IPR and UAR 2025 ScanMAT) who made the Raman measurements from Ji's work in 2007, for fruitful discussions about confocal Raman spectroscopy and M. Nivard for performing the nano-indentation tests.

\section*{References}
\bibliographystyle{model3-num-names-vk}
\begin{footnotesize}
    \bibliography{Publis_Verre_2022}
\end{footnotesize}


\clearpage

\clearpage

\clearpage

\clearpage

\clearpage

\clearpage

\clearpage

\clearpage

\clearpage

\clearpage
\begin {figure}[ht]
\centering
\includegraphics[angle=90,width=.75\textwidth]{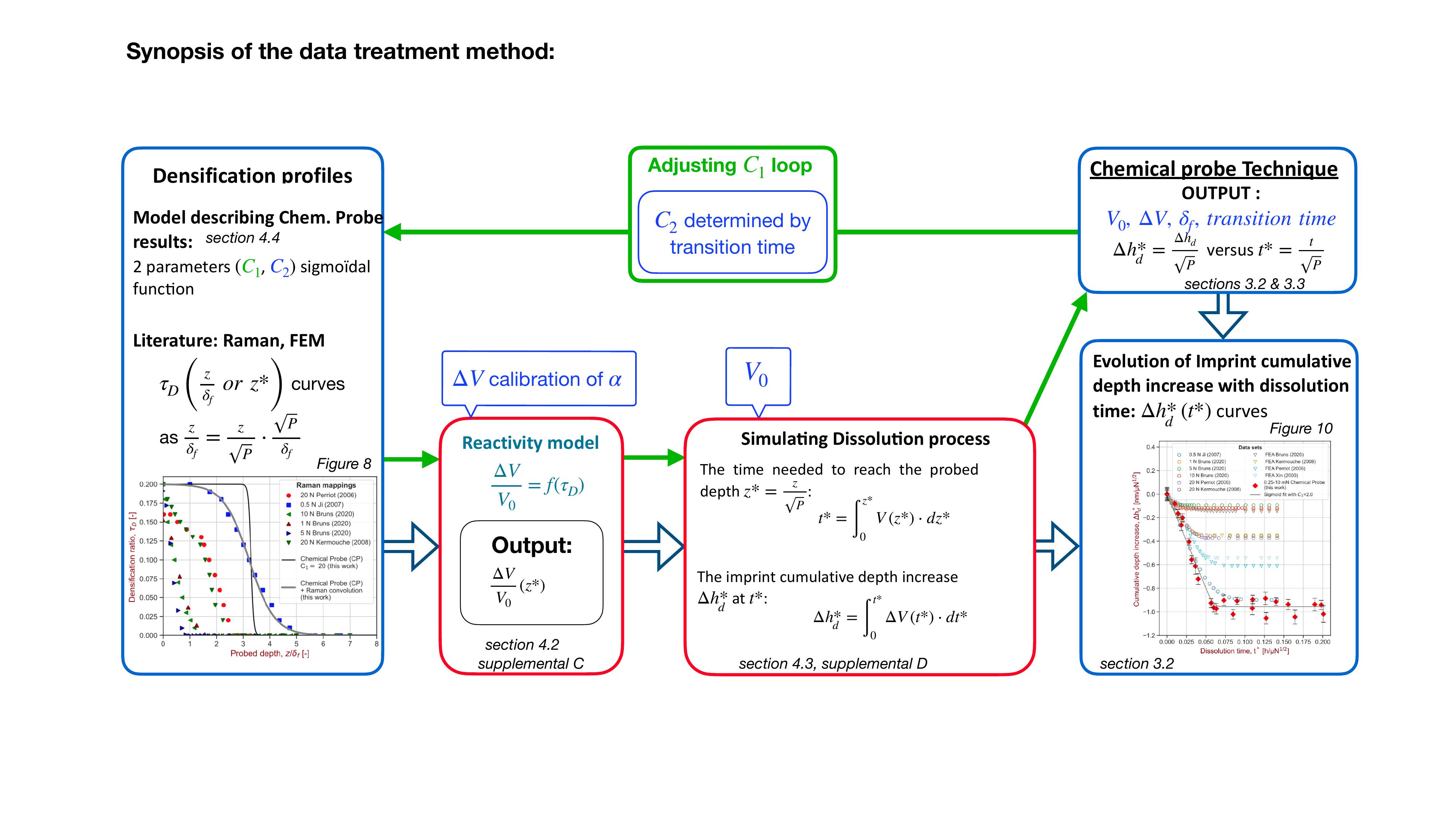}
\caption{Synopsis of data analysis between \Cref{fig:profiles_lit} and \Cref{fig:dh_all}.}
\label{fig:synopsis}
\end{figure}

\clearpage

\end{document}